\documentclass[twocolumn,showpacs,preprintnumbers,amsmath,amssymb,superscriptaddress]{revtex4} 
\usepackage{epsfig,bm}
\usepackage{amsmath,amssymb}
\usepackage{color}
\newcommand{\beag}{\mbox{$^{8}{\rm Be}(\alpha,\gamma)^{12}$C}}
\begin{document}
\title{Barrier penetration effects in the triple-alpha reaction at low energies } 
\author{Y. Suzuki}
\email{suzuki@nt.sc.niigata-u.ac.jp}
\affiliation{Department of Physics, Niigata University, Niigata
950-2181, Japan} 
\affiliation{RIKEN Nishina Center, Wako 351-0198, Japan} 
\author{P. Descouvemont}
\email{pdesc@ulb.ac.be}
\affiliation{Physique Nucl\'eaire Th\'eorique et Physique 
Math\'ematique, C.P.229, Universit\'e Libre de Bruxelles (ULB), B-1050 Brussels, Belgium}
\pacs{26.20.Fj, 25.40.Lw,21.45.-v}

\begin{abstract}
We investigate the triple-alpha reaction  at low energies, by
assuming a direct process. The Coulomb potential of 
three $\alpha$ particles is examined carefully. The three-body continuum 
wave functions are generated by calculating an adiabatic potential
barrier. We discuss the influence of the $\alpha\alpha$ potential, and compare our
reaction rates with the literature.
The reaction rate at $T=0.01$\,GK is 
about 10$^3$ times larger than that of NACRE.    
\end{abstract}
\maketitle

\section{Introduction}
The triple-$\alpha$ process plays an important role in stellar physics, since it triggers
helium burning in stars. Owing to the absence of stable isotopes of mass 5 and 8, it represents the only
way to synthesize $^{12}$C in stars. At typical helium burning temperatures, the triple-$\alpha$ process is assumed to be sequential \cite{Ho54}. Two $\alpha$ particles are in equilibrium with the unstable
$^8$Be isotope, which then captures a third $\alpha$. The second step, the $\beag$ reaction, is
strongly influenced by the $^{12}$C $0^+_2$ state ($E_x=7.65$ MeV), located 0.29 MeV
above the $\alpha+^8$Be threshold. This resonance was predicted by Hoyle \cite{Ho54} and
found experimentally later \cite{DPW53}. It is known, in the nuclear astrophysics community, as the
Hoyle state. The properties (energy, alpha and gamma widths) of this resonance are well
known experimentally.

The calculation of the triple-$\alpha$ reaction rate is, however, a subject of intense debate.
First calculations \cite{NTM85,LWT86,nacre} were based on the hypothesis of a sequential process, 
where the resonant nature of $^8$Be and of the $^{12}$C$(0^+_2$) resonance are accounted for
by a Breit-Wigner (BW) approximation. For narrow resonances ($\Gamma_{\alpha}=6$ eV in $^8$Be, and
$\Gamma_{\alpha}=8.5$ eV in the Hoyle state), the BW formalism is expected to
provide a fair approximation of the phase shifts and cross sections.

This simple approach was recently challenged by Ogata {\it et al.}~\cite{ogata09} who use the continuum-discretized coupled-channels (CDCC) method~\cite{YIK86}. In the CDCC theory, the $^8$Be continuum
is simulated by approximate, square-integrable wave functions. The CDCC method provides a consistent
way to include the $\alpha+\alpha$ continuum in the calculation of the triple-$\alpha$
reaction rate. However, as the involved resonances are narrow, the BW approximation is expected to be reliable. CDCC results should
therefore not be very different from results obtained previously in the literature. In fact, Ogata {\it et al.} 
found reaction rates much higher than those of NACRE (20 orders of magnitude at low temperatures).
At typical helium burning temperatures, the CDCC rates, when included in stellar models, are
incompatible with observation~\cite{DP09}.

The unexpectedly large CDCC results have triggered several theoretical studies, in different
models. In particular the sequential picture was questioned by 
Garrido {\it et al.}~\cite{garrido11} who suggest
to extend the standard BW approximation
to the three-$\alpha$ capture, which does not go
through the $^8$Be resonance. This simple direct model was then improved by 
Nguyen {\it et al.}~\cite{nguyen12} who use the hyperspherical formalism, associated with the $R$-matrix theory (HHR), to
determine solutions of the $\alpha+\alpha+\alpha$ scattering problem. These approaches
essentially confirm the NACRE reaction rate above $T\approx 0.1$\,GK, where helium burning occurs.
The large CDCC results cannot be explained. More recently, Ishikawa~\cite{ishikawa13}
investigated the triple-$\alpha$
problem in the Faddeev formalism. The results are essentially in agreement with NACRE, and
much smaller than the recent three-body calculations.

Considering that the CDCC reaction rates above $T=0.1$\,GK are obviously inconsistent with observation, all other models agree in
this temperature region, relevant for astrophysics. At lower temperatures, however, significant
differences still exist between the various approaches. If the capture cross sections at these
low energies are of minor importance in astrophysics, they raise interesting questions for nuclear
physics aspects. The triple-$\alpha$ reaction rate offers a unique opportunity to investigate
further three-body problems, in particular with three charged particles.

In this work, we address qualitatively the $3\alpha$ system at low energies assuming a direct process, 
and we focus 
on general properties of the Coulomb interaction. 
A specific purpose is an attempt at understanding  
the reason for the differences of several orders of magnitude. 
In Sec.~\ref{reaction.rate.formula} we summarize  
the basic formulas to calculate the triple-$\alpha$ 
reaction rate according to the sequential and direct 
capture processes. We discuss in 
Sec.~\ref{Coulomb.three-body} general features of the 
Coulomb barrier for 3 $\alpha$ particles.  
The method used here is  
applicable for any 
Coulomb three-body problem. An adiabatic 
potential including the nuclear contribution is discussed 
in Sec.~\ref{barrier}. The triple-$\alpha$ reaction rate 
is estimated assuming a simple $2^+$ wave function of 
$^{12}$C in Sec.~\ref{insight}. Conclusion is drawn in Sec.~\ref{conclusion}. There are four appendices. We 
discuss totally symmetric hyperspherical harmonics (HH) 
in Appendix~\ref{app.symm.pol}, calculate  
the Coulomb matrix element  
in the hyperspherical coordinate in Appendix~\ref{coulomb.m.e}, 
examine the convergence of the 
HH expansion in Appendix~\ref{hhe}, and analyze  
partial wave contents produced by symmetrization in Appendix~\ref{pwc}.   

\section{Triple-$\alpha$ reaction rate }
\label{reaction.rate.formula}
\subsection{Sequential capture}
Let us consider a process $2+3+4 \to 0+1$, where e.g., 2 stands for 
a nucleus with mass $A_2$ in the mass unit $m$. All masses 
of particles 2, 3 and 4 are assumed to be the same. Let $\langle 234
\rangle _{seq}$ denote the sequential reaction rate $R_{234}(E)$.
In this approximation, the triple-$\alpha$ capture is assumed to proceed in two steps. According
to Refs. \cite{NTM85,LWT86}, the reaction rate is given by
\begin{align}
&\langle 234
\rangle _{seq}= 3
\left( \frac{8\pi\hbar}{\mu_{\alpha\alpha}^2} \right)
\left( \frac{\mu_{\alpha\alpha}}{2\pi k_{\rm B} T} \right)^{3/2} \notag \\
& \times
\int_0^{\infty}
\frac{\sigma_{\alpha\alpha}(E)}{\Gamma_{\alpha}(^8{\rm Be},E)}
\exp\big(-{\textstyle{\frac{E}{k_{\rm B} T}}}\big)  \langle \sigma v \rangle^{\alpha ^8{\rm Be}}
\, E \, dE,
\end{align}
where $\mu_{\alpha\alpha}$ is the reduced mass
of the $\alpha$ + $\alpha$ system, and $E$ is the energy with
respect to the $\alpha$ + $\alpha$ threshold.
The elastic cross section of $\alpha$ + $\alpha$ scattering is
given by a BW approximation, where the width of the $^8$Be ground state is energy dependent.

The $ \langle \sigma v \rangle^{\alpha^8{\rm Be}}$ rate assumes
that $^8$Be has been formed at an energy $E$ different from $E_{^8 \rm Be}$,
and that it is bound.
This rate is given by
\begin{align}
&\langle \sigma v \rangle^{\alpha ^8{\rm Be}} =
\frac{8\pi}{\mu_{\alpha ^8{\rm Be}}^2}
\left( \frac{\mu_{\alpha ^8{\rm Be}}}{2\pi k_{\rm B} T} \right)^{3/2} \notag \\
&\times \int_0^{\infty} \sigma_{\alpha ^8{\rm Be}}(E';E)
\exp\big(-{\textstyle{\frac{E'}{k_{\rm B} T}}}\big)\ E'\, dE',
\end{align}
where $\mu_{\alpha ^8{\rm Be}}$ is the reduced mass of the $\alpha$ + $^8$Be
system, and $E'$ is the energy with respect
to its threshold (which varies with the formation energy $E$). Cross section
$\sigma_{\alpha ^8{\rm Be}}(E';E)$ corresponds to the $\beag$ reaction, where $E'$ is
the $\alpha+^8$Be energy.
This formalism has been used in the NACRE compilation for the calculation of
the triple-$\alpha$ process \cite{nacre}.

\subsection{Three-body capture}
We present here a brief overview of the three-body capture. More detail can be found in
Refs.~\cite{fowler67,garrido11}.
Let $\langle 234
\rangle ^{3b}$ denote the reaction rate $R_{234}(E)$ averaged over 
the energy distribution
\begin{align}
\langle 234 \rangle  ^{3b}= \int dE R_{234}(E)
\frac{1}{2(k_BT)^3}E^2\exp\big(-{\textstyle{\frac{E}{k_BT}}}\big).
\label{tnr1}
\end{align}
Similarly the inverse reaction process $0+1 \to 2+3+4$ has the
thermonuclear reaction rate given by
\begin{align}
\langle 01 \rangle &= \langle \sigma v \rangle_{01} \notag \\
&=\int dE' \sigma_{01}(E')v'
 \frac{2}{\sqrt{\pi}(k_BT)^{3/2}}\sqrt{E'}\exp(-\textstyle{\frac{E'}{k_BT}})\notag
 \\
&=\frac{2\sqrt{2}}{\sqrt{\pi \mu_{01}}(k_BT)^{3/2}}\int dE' \sigma_{01}(E')
 E'\exp(-\textstyle{\frac{E'}{k_BT}}),
\label{tnr2}
\end{align}
where $E'=E-Q$ with the $Q$ value for the reaction $0+1 \to 2+3+4$. 
In the case of the triple-$\alpha$ reaction, the $Q$ value is $Q=-2.836$\,
MeV. 
We want to relate $R_{234}(E)$ to $\sigma_{01}(E')$.

The use of the detailed balance or reciprocity relation leads to 
\begin{align}
\frac{\langle 234 \rangle}{\langle 01 \rangle} 
&=\frac{g_0g_1}{g_2g_3g_4}
\left(\frac{A_0A_1}{A_2A_3A_4}\right)^{3/2}\notag \\
&\times \left(\frac{2\pi\hbar^2}{mk_BT}\right)^{3/2}\frac{1+\Delta_{234}}{1+\delta_{01}}\exp(-\textstyle{\frac{Q}{k_BT}}).  
\label{tnr3}
\end{align}
The validity of Eq.~(\ref{tnr3}) relies on the time-reversal invariance of 
the Hamiltonian. It also implies that the first-order perturbation theory  can be used.

In the case of $a+b+c \to A+\gamma$ process, $\sigma_{01}$ denotes the
photoabsorption cross section $\sigma_{\gamma}$ with $E_{\gamma}=E-Q$
for $A+\gamma \to a+b+c$, and 
the above relation leads to the desired expression  
\begin{align}
R_{abc}(E)&=(1+\Delta_{abc})\frac{2g_A}{g_ag_bg_c} 
\frac{8\pi \hbar^3
 }{(\mu_{ab}\mu_{ab,c})^{3/2}c^2}\notag \\
&\times \left(\frac{E_{\gamma}}{E}\right)^2
\sigma_{\gamma}(E_{\gamma}).
\label{r(E)}
\end{align}
See Refs.~\cite{fowler67,garrido11} for the notations.  
If
the photoabsorption occurs by an electric multipole $E\lambda$, its 
photoabsorption cross section 
can be expressed in terms of the strength function 
$S_{E\lambda}(E_{\gamma})$ by 
\begin{align}
\sigma_{\gamma}(E_{\gamma})
=\frac{(2\pi)^3(\lambda+1)}{\lambda((2\lambda+1)!!)^2}
\left(\frac{E_{\gamma}}{\hbar
 c}\right)^{2\lambda-1}S_{E\lambda}(E_{\gamma}), 
\label{photoabs}
\end{align}
where 
\begin{align}
&S_{E\lambda}(E_{\gamma})\notag \\
& =\frac{1}{2J_i+1}{\cal S}_{f}|\langle \Psi_f||{\cal
 M}_{E\lambda}||\Psi_i \rangle|^2\delta(E_f-E_i-E_{\gamma}). 
\label{strength.fn}
\end{align}
Here ${\cal M}_{E\lambda}$ is the $E\lambda$ operator, $J_i$ is the angular momentum of the 
initial state and notation ${\cal S}_{f}$ includes the integration over the final state
energy $E_f$ as well as the summation over the other quantum numbers that specify the final state. 

\section{Triple-$\alpha$ Coulomb potential}
\label{Coulomb.three-body}
\subsection{Hyperspherical coordinates}
The Coulomb potential 
for a system including three charged particles is most 
transparently treated in the hyperspherical coordinates. 
Here we also express those operators that are relevant to the 
triple-$\alpha$ reaction in terms of the hyperspherical coordinates.

Let $\bm{r}_i$ denote the coordinate of $i$th 
$\alpha$ particle. The intrinsic motion of 3\,$\alpha$ system is described 
with two relative coordinates
\begin{align}
\bm{x}_1=\frac{1}{\sqrt{2}}(\bm{r}_1-\bm{r}_2),\quad
\bm{x}_2=\sqrt{\frac{2}{3}}\biggl(\frac{\bm{r}_1+\bm{r}_2}{2}-\bm{r}_3\biggr). 
\label{jacobi.coord}
\end{align}
Other Jacobi coordinate sets $\bm y_1, \bm y_2$ and 
$\bm z_1, \bm z_2$ are respectively obtained by cyclic 
permutations $(1,2,3)$ and $(1,3,2)$ from $\bm x_1, \bm x_2$. 
The center of mass coordinate is denoted as   
$\bm x_3=(\bm r_1+\bm r_2+\bm r_3)/3$.

The hyperradius $\rho$ and five angular coordinates 
$\Omega_x$ are used in 
the HH method (see Ref.~\cite{danilin} for details). 
Four angle coordinates of $\Omega_x$ come from the angular coordinates
$\hat{\bm x}_1=\bm x_1/x_1$ and $\hat{\bm x}_2=\bm x_2/x_2$, and the 
hyperangle $\alpha$ ($0 \leq \alpha \leq \pi/2$) is defined by 
\begin{align}
x_1=\rho \cos\alpha,\quad x_2=\rho \sin\alpha. 
\label{hyperangle}
\end{align}
The hyperradius $\rho$ is expressed in various ways: 
\begin{align}
\rho^2&=x_1^2+x_2^2=\frac{1}{3}\sum_{j>i=1}^3(\bm r_i-\bm r_j)^2 \notag \\
&=\sum_{i=1}^3(\bm r_i-\bm x_3)^2\equiv 3{\cal M}_{00}.
\label{hyperradius}
\end{align}
Here ${\cal M}_{00}$ is the operator for the mean square radius. Note the difference in 
the definition of $\rho$ in the literature. Our $\rho$ is a half of 
the hyperradius of Ref.~\cite{nguyen12} and 1/$\sqrt{2}$ 
of the hyperradius of Ref.~\cite{ishikawa13}.  
Note also that the volume element for the 
six-dimensional integral  
is expressed as 
$d{\bm x}_1d{\bm x}_2=\rho^5\cos^2\alpha \sin^2\alpha \,
d\rho \,d\alpha \, d{\hat{\bm x}_1}d{\hat{\bm x}_2}$.

The kinetic energy $T$ of the 3\,$\alpha$ system, with the center of mass 
kinetic energy being subtracted, is expressed as  
\begin{align}
T&=-\frac{\hbar^2}{2m_{\alpha}}\left(\frac{\partial^2}{\partial \bm
 x_1^2}+\frac{\partial^2}{\partial \bm x_2^2}\right)\notag \\
&=-\frac{\hbar^2}{2m_{\alpha}}\left(\frac{\partial^2}{\partial \rho^2}
+\frac{5}{\rho}\frac{\partial}{\partial \rho}-\frac{1}{\rho^2}{\cal K}^2
\right),  
\label{kinetic.op}
\end{align}
where the hypermomentum operator ${\cal K}^2$ is given by  
\begin{align}
{\cal K}^2=-\frac{\partial^2}{\partial \alpha^2}-4\cot 2\alpha 
\frac{\partial}{\partial \alpha}+\frac{1}{\cos^2 \alpha}
{\bm \ell}_1^2+\frac{1}{\sin^2 \alpha}{\bm \ell}_2^2.
\label{hyp.momentum}
\end{align}
Here ${\bm \ell}_1$ and ${\bm \ell}_2$ are the angular momenta 
corresponding to the coordinates ${\bm x}_1$ and 
${\bm x}_2$, respectively.

The $E2$ operator is expressed as 
\begin{align}
{\cal M}_{E2m}&=2e\sum_{i=1}^3{\cal Y}_{2m}(\bm r_i-\bm x_3) \notag \\
&=2e[{\cal Y}_{2m}(\bm
 x_1)+{\cal Y}_{2m}(\bm x_2)] ,
\label{E2op}
\end{align}
with ${\cal Y}_{\ell m}(\bm r)=r^{\ell}Y_{\ell}^{m}(\hat{\bm r})$. 
Because $\rho^2, {\cal K}^2$ and ${\cal M}_{E2m}$ 
are all symmetric operators,  the
$\bm x$ coordinates can be replaced by the
$\bm y$ or $\bm z$ coordinates, e.g., $\rho^2=y_1^2+y_2^2$ and 
${\cal M}_{E2m}=2e[{\cal Y}_{2m}(\bm z_1)+{\cal Y}_{2m}(\bm z_2)]$.

\subsection{Triple-$\alpha$ Coulomb barrier}
\label{triple-alpha.Coulomb}

The Coulomb potential for 3\,$\alpha$ particles 
can  be treated in the HH method. Particularly we determine the Coulomb potential that gives 
the most enhanced triple-$\alpha$ reaction rate 
at low energies. In other words, the Coulomb potential should be 
as low as possible for 3\,$\alpha$ particles.  

The Coulomb potential for 3\,$\alpha$ particles 
is expressed in hyperspherical coordinates as  
\begin{align}
V_C=\sum_{j>i=1}^3\frac{4e^2}{|\bm r_i-\bm
 r_j|}=\frac{4e^2}{\rho}q(\Omega_x)
 \label{coulomb.3alpha}
\end{align}
with a `charge factor' operator 
\begin{align}
q(\Omega_x)&=\frac{1}{\sqrt{2}}\Big\{\frac{1}{\cos \alpha}+\frac{1}
{|-\frac{1}{2}\cos \alpha \, \hat{\bm x}_1+
\frac{\sqrt{3}}{2}\sin \alpha \, \hat{\bm x}_2|}\notag \\
& \qquad \ \ 
+\frac{1}{|-\frac{1}{2}\cos \alpha \, \hat{\bm x}_1-
\frac{\sqrt{3}}{2}\sin \alpha \, \hat{\bm x}_2|}\Big\}.
\label{qfactor.def}
\end{align}
As $q(\Omega_x)$ is
symmetric with respect to the coordinate transformation, 
we may omit the suffix $x$ of $\Omega_x$.

Let us consider the eigenvalue problem 
\begin{align}
q(\Omega)F_q(\Omega)=qF_q(\Omega).
\label{q.eig.prob}
\end{align}
The eigenfunction $F_q(\Omega)$ must be symmetric with respect to 
the permutations of particles 1,2,3.
The eigenvalue problem is solved using a complete set of 
the HH basis, ${F}_{KLM}^{{\ell}_1{\ell}_2}(\Omega_x)$. 
Let $\{{F}_{KLM}^{\gamma}\}$ 
denote an orthonormal set of symmetric functions constructed  from the HH basis, where $\gamma$ is a
suitable label to enumerate the symmetric functions (see Appendix~\ref{app.symm.pol} for details). 
A solution $F_q(\Omega)$ is expanded in the set 
$\{{F}_{KLM}^{\gamma}\}$. 
Unfortunately $q(\Omega)$ has non-vanishing
matrix elements for $K\neq K'$, which is in fact the origin 
that makes a complete treatment of the Coulomb three-body problem extremely hard. A method of calculating the matrix element of $q(\Omega)$ 
is presented in Appendix~\ref{coulomb.m.e}.

Figure~\ref{qeig.value} 
displays the spectrum of eigenvalues for $L=0$ case 
as a function of 
$K_{\rm max}$, a maximum $K$ value included in the diagonalization. Table~\ref{q.data} lists 
$N$ (the basis dimension at $K_{\rm max}$), $q_{\rm min}$ and $q_{\rm max}$, minimum and maximum 
eigenvalues of $q(\Omega)$ as well as $\langle q \rangle$, 
an average of the 
expectation values of $q(\Omega)$, i.e. the average of the 
eigenvalues 
\begin{align}
\langle q \rangle 
=\frac{1}{N}\sum_{\gamma=1}^N \langle {
 F}_{KLM}^{\gamma}|q(\Omega)|{F}_{KLM}^{\gamma}\rangle
=\frac{1}{N}{\rm Tr}\,q(\Omega).
\end{align}
The $\langle q\rangle$ value increases very
slowly as $K_{\rm max}$ increases. It starts from 3.601 at $K_{\rm max}=0$
and reaches about 4.25 at $K_{\rm max}=70$. It is remarkable that $q_{\rm min}$ approaches 3 and 
does not become 
smaller than 3.  On the other hand, $q_{\rm max}$ increases monotonically as a function of 
$K_{\rm max}$. 

As seen above, the spectrum of eigenvalues 
varies as a function of 
$K_{\rm max}$. This is 
a consequence of possible arrangements that
3$\alpha$ particles can take for a given $K_{\rm max}$. 
With increasing $K_{\rm max}$ not only  
$q_{\rm min}$ approaches 3 but also an increasing number of eigenfunctions have 
$q$ values very close to 3. Classically we expect  
$q=3$ assuming  
a regular triangle configuration of 3\,$\alpha$ particles because then $\rho$ is equal to its side  
length as indicated by Eq.~(\ref{hyperradius}) and its Coulomb
potential, $3\times 4e^2/\rho$, gives $q=3$. The value of 
$q_{\rm max}$ is, however, isolated from the other eigenvalues. It approximately follows a straight line, 
$q_{\rm max}\approx 0.35 K_{\rm max}+2.8$ for $K_{\rm max} \geq 4$. 

\begin{figure}
\begin{center}
\epsfig{file=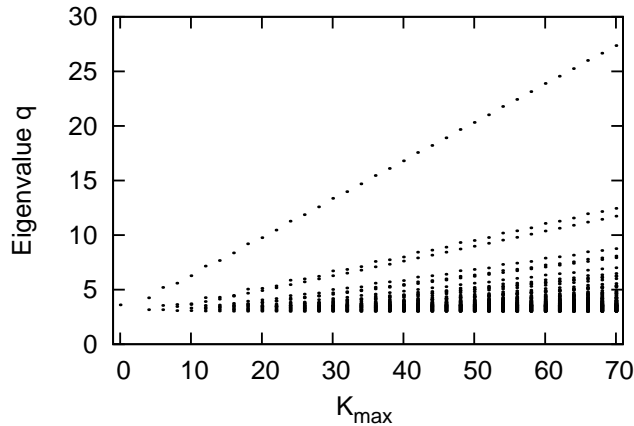,scale=1.4}
\caption{
Eigenvalues of $q$ for $3\alpha$ Coulomb potential as a function of $K_{\rm max}$.}
\label{qeig.value}
\end{center}
\end{figure}

\begin{table}
\begin{center}
\caption{Properties of the eigenvalues of 3$\alpha$ charge factor~(\ref{qfactor.def}) as a 
 function of $K_{\rm max}$. }
\label{q.data}
\begin{tabular}{ccccccccc}
\hline\hline
$K_{\rm max}$& 0 &10 &20 & 30 &40 &50 &60 & 70   \\
\hline
$N$& 1 &5 &14 & 27 &44 &65 &91 & 120   \\
$q_{\rm min}$&3.601&3.076&3.021&3.012&3.006&3.005&3.003&3.002\\
$q_{\rm max}$&3.601&6.268&9.770&13.37&16.80&20.33&23.89&27.37  \\
$\langle q\rangle$&3.601&4.000&4.140&4.214&4.216&4.236&4.247&4.248  \\
\hline\hline
\end{tabular}
\end{center}
\end{table}

The Coulomb potential is now represented as 
$V_C(\rho)=4e^2q/\rho$ with use of the eigenvalue $q$.  
In the representation that diagonalizes $q(\Omega)$ 
the Coulomb potential has no coupling 
at all.  
The value of $q_{\rm min}$ is of 
particular interest in the low energy triple-$\alpha$ reaction 
because it leads to a minimum Coulomb barrier.

\section{Adiabatic potential barrier}
\label{barrier}

The initial state $\Psi_i$ of Eq.~(\ref{strength.fn}) is
confined in a small region around the center of mass of the 3\,$\alpha$
particles. However, the final continuum state at very low energy has to penetrate through a thick  barrier to reach the asymptotic region 
where its normalization is preset. It is crucially important to evaluate properly  
the barrier for 3\,$\alpha$ particles in order to obtain the 
photoabsorption cross section at low energies. The barrier 
may be calculated in the adiabatic hyperspherical 
expansion method~\cite{fedorov96,fedotov04}, where the potential 
energy acting among $\alpha$ particles is 
diagonalyzed on the hypersphere with a radius $\rho$. Repeating this 
calculation for a number of $\rho$ values, one obtains an adiabatic potential. 
This may be akin in spirit 
to what is done for the Coulomb potential in Sect.~\ref{triple-alpha.Coulomb}. 
Contrary to the Coulomb case where   
the $\rho$-dependence is trivially known,  
one has to repeat the diagonalization at each $\rho$.  

Since the HH expansion is known to converge  
slowly as discussed in Appendix~\ref{hhe}, one may take other 
trial function, $\Psi=\sum_{\xi}C_{\xi}\Phi(\xi)$, where $\Phi(\xi)$ is a 
basis function depending on some parameters $\xi$, The barrier at $\rho$ can  
be obtained by minimizing the energy $\langle \Psi|H|\Psi
\rangle/\langle \Psi|\Psi \rangle$ with respect to $C_{\xi}$ under the 
constraint $\langle \rho^2 \rangle=\rho^2$.  
Here we adopt a product of Gauss wave packets specified by two `generator 
coordinates' $\bm s_1$ and $\bm s_2$
\begin{align}
\Phi(\bm s_1, \bm s_2, \bm
 x)&=\left(\frac{\beta}{\pi}\right)^{3/4}\exp(-{\textstyle{\frac{1}{2}\beta(\bm
 x_1-\bm s_1)^2}})\notag \\
&\times \left(\frac{\beta}{\pi}\right)^{3/4}\exp(-{\textstyle{\frac{1}{2}\beta(\bm x_2-\bm s_2)^2}}). 
\label{gcmwf}
\end{align}
The value of $\beta$ is related to the mean square radius 
of $\alpha$ particle. It  
is chosen as $\beta=4\times 0.52$\,fm$^{-2}$ based on  
the $(0s)^4$ harmonic-oscillator shell-model wave function for  $\alpha$ 
particle~\cite{matsumura04}. We symmetrize the wave function 
$\Phi(\bm s_1, \bm s_2, \bm x)$ to calculate the potential energy matrix element. 
The expectation value of the potential energy, 
$V(\bm s_1, \bm s_2)$,  
is an estimate of the barrier corresponding to 
the geometric arrangement specified by $\bm s_1$ and $\bm s_2$. 

In this study we do not minimize the energy but approximately obtain the barrier 
by averaging $V(\bm s_1, \bm s_2)$ over $\bm s_1$ and 
$\bm s_2$ with the constraint of $s_1^2+s_2^2=s^2=\rho^2$:
\begin{align}
V(\rho)=\frac{1}{\rho^5}\int d{\bm s}_1\int d{\bm s}_s 
V(\bm s_1, \bm s_2)\, w(\Omega_s)\delta(s-\rho)
\label{ave.pot}
\end{align} 
with a weight function $w(\Omega_s)$ that satisfies the normalization 
\begin{align}
\int d\Omega_s w(\Omega_s) =1.
\label{norm.condition}
\end{align}
The  $1/\rho^5$ factor of Eq.~(\ref{ave.pot}) arises because of 
the property  
$\int d{\bm s}_1\int d{\bm s}_s \, w(\Omega_s)\delta(s-\rho)
=\int ds\, s^5 \int d\Omega_s w(\Omega_s)\delta(s-\rho)
=\rho^5$. 
Any $w(\Omega_s)$ that meets Eq.~(\ref{norm.condition}) is expressed as 
\begin{align}
w(\Omega_s)=\sum_{K\ell}C_{K\ell}{F}^{\ell \ell}_{K00}(\Omega_s)
\end{align}
with the constraint $C_{00}={F}^{00}_{000}(\Omega_s)=\pi^{-3/2}$.

We choose $C_{K\ell}=C_{00}\delta_{K,0}\delta_{\ell,0}$ to 
project out only $K=0$ barrier, which 
leads to the adiabatic potential 
\begin{align}
V(\rho)=\frac{1}{\pi^3\rho^5}\int d{\bm s}_1\int d{\bm s}_s 
V(\bm s_1, \bm s_2)\, \delta(s-\rho).
\label{ave.pot1}
\end{align}
Note that the adiabatic potential of this choice approaches  
$4e^2\times 3.601/\rho$ for large $\rho$, where only the Coulomb 
potential contributes to the barrier. The value of 3.601 is the 
eigenvalue of $q$ at $K=0$. We modify the Coulomb contribution to its minimum 
at large $\rho$ in calculating the photoabsorption cross section.
 
The kinetic energy~(\ref{kinetic.op}) contains the centrifugal potential. 
Its form is apparent for $\psi=\rho^{5/2}\Psi_f$~\cite{danilin}, and its 
contribution to the adiabatic potential reads 
\begin{align}
V_{CF}(\rho)&=\frac{\hbar^2}{2m_{\alpha}\rho^2}
\int d{\Omega}_s  [{\cal K}^2(\Omega_s) +{\textstyle{\frac{15}{4}}}] w(\Omega_s)\notag \\
&=\frac{\hbar^2}{2m_{\alpha}\rho^2}\frac{15}{4}.
\end{align}
Thus the centrifugal potential also becomes a minimum. As the result our 
adiabatic potential barrier turns out to be lowest regarding the 
Coulomb and centrifugal potentials. In this case the centrifugal potential becomes less than 1\% of 
the Coulomb potential only when $\rho$ is larger than 115\,fm.

\begin{figure}
\begin{center}
\epsfig{file=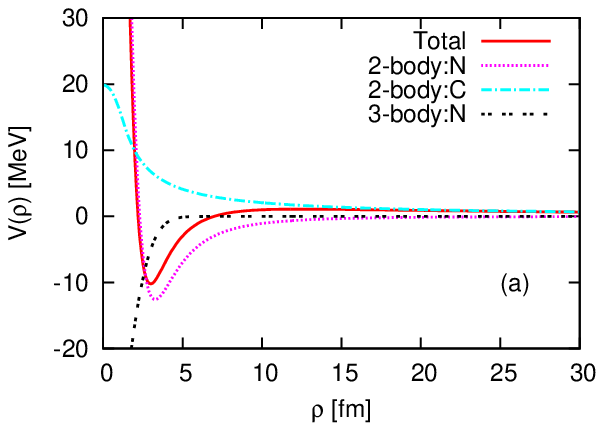,scale=1.4}
\epsfig{file=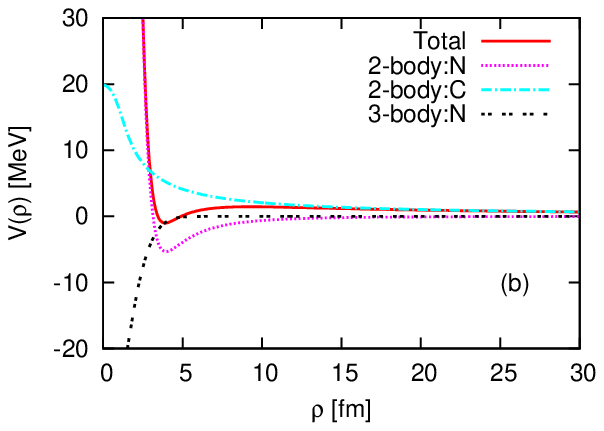,scale=1.4}
\caption{
(Color online) The adiabatic potential for AB(A$^{\prime}$) (a) and
 AB(D) (b) potentials  including 3-body potential. The potential
 parameters are taken from Ref.~\cite{ishikawa13}.} 
\label{abpotential}
\end{center}
\end{figure}

Panels (a) and (b) of Fig.~\ref{abpotential} display  
the adiabatic potentials calculated with Ali-Bodmer A$^{\prime}$ (AB(A$^{\prime}$)) and Ali-Bodmer D (AB(D)) $\alpha \alpha$ 
potentials~\cite{ali66,fedorov96}, respectively. 
The mass of $\alpha$ particle is  
${\hbar^2}/{m_{\alpha}}=10.5254\,{\rm MeV\,fm^2}$, and 
the charge constant is $e^2=1.43996\,{\rm MeV\,fm}$. 
The contributions 
from the $\alpha \alpha$ nuclear and Coulomb potentials 
as well as the three-body force are also shown. 
All of the potential parameters are the same as 
those of Ref.~\cite{ishikawa13}. The AB(A$^{\prime}$) potential 
is used also in Ref.~\cite{nguyen12}, but its  
three-body force is different from the present one. 
The barrier peak is 1.1\,MeV at $\rho=12.4$\,fm for AB(A$^{\prime}$) and 1.5\,MeV at $\rho=9.3$\,fm for AB(D), respectively.  
Because the AB(D) potential contains stronger repulsion at short distances, the minimum of the adiabatic potential is rather shallow and located 
at distance larger than that of AB(A$^{\prime}$) potential.  
The adiabatic potential 
of AB(A$^{\prime}$) is deep enough 
to produce a resonance. The energy and width of the resonance 
are 0.702\,MeV and 6\,keV, so that its energy is too high 
and its width is too wide to be compared to those  
 (0.379\,MeV, 8.5\,eV) of the Hoyle state. 
Note that each piece of the Hamiltonian contributes  differently in its distance of reach.  
The $\alpha \alpha$ nuclear contribution decreases 
rather slowly as a function of $\rho$.  
In the case of AB(D) potential, it becomes 1\% of the Coulomb contribution 
only when $\rho$ is larger than 50 fm.

To see the effect of the symmetrization, we also calculate 
the adiabatic potential using the wave function~(\ref{gcmwf}) 
itself. Most noteworthy is that the Coulomb barrier evaluated by ignoring the boson symmetry is smaller by about 7\% 
for any $\rho$ than the one calculated with the 
symmetrized wave function. 
For example, the non-symmetrized Coulomb barrier approaches 
$4e^2\times 3.35/\rho$ for large $\rho$ instead of 
$4e^2\times 3.60/\rho$.

\section{Calculation of triple-$\alpha$ reaction rates at low energies}
\label{insight}
\subsection{Wave functions}
The triple-$\alpha$ reaction rate~(\ref{r(E)}) can be calculated  
from the photoabsorption cross section 
$\sigma_{\gamma}(E_{\gamma})$ of Eq.~(\ref{photoabs}) for 
$E2$. The result is  
\begin{align}
R_{\alpha \alpha \alpha}(E)&=1440\sqrt{3}\pi\frac{\hbar^3}{m_{\alpha}^3c^2}
\left(\frac{E_{\gamma}}{E}\right)^2\sigma_{\gamma}(E_{\gamma}),
\end{align}
where $E_{\gamma}=E+2.836$ in MeV.

To obtain $\sigma_{\gamma}$, we take a simple model for the initial and final states. 
The $2^+$ initial bound state is assumed to be 
\begin{align}
\Psi_{2M} ={\cal N}\exp(-{\textstyle{\frac{1}{2}a\rho^2}})
({\cal Y}_{2M}(\bm x_1)+{\cal Y}_{2M}(\bm x_2)),
\label{init.state}
\end{align}
where ${\cal N}=\sqrt{8a^5/15\pi^2}$ is the normalization constant. This wave function 
satisfies the boson symmetry. The parameter $a$ is related to the root-mean-square (rms) 
radius of the 2$^+$ state of $^{12}$C. The mean square radius 
$\langle \rho^2\rangle$ of the 3\,$\alpha$ system 
is related to that of $^{12}$C, $\langle r^2_C \rangle$,  
by the following relation 
\begin{align}
\langle r^2_C \rangle=\langle r^2_{\alpha} \rangle + 
\langle {\cal M}_{00} \rangle=
 \langle r^2_{\alpha} \rangle +\frac{1}{3}\langle \rho^2 \rangle,
\label{rms.relation}
\end{align}
where 
$\langle r^2_{\alpha} \rangle$ is the mean square radius of $\alpha$ particle.  The value of $\langle \rho^2\rangle$ calculated with 
Eq.~(\ref{init.state}) is  
$\langle \rho^2\rangle=5/a$.  
We adopt $\sqrt{\langle r^2_C \rangle} \approx 2.45$\,fm based on theoretical 
calculations~\cite{fmd}, which leads to a choice of   
$a=0.43$\,fm$^{-2}$. 

To check the reliability of the $2^+$ wave function~(\ref{init.state}) for  evaluating 
the $E2$ matrix element, we determine the quadrupole moment of the 
$2^+$ state. 
The intrinsic quadrupole moment, 
$Q_0=\sqrt{16\pi/5} \langle \Psi_{22}|{\cal M}_{E20}|\Psi_{22}\rangle $, is obtained as  $Q_0=-(4/a)\,e=-9.3\,e$\,fm$^2$, which is compared to the experimental value,  
$-22\pm10\,e$\,fm$^2$~\cite{vermeer83}. Our model wave function appears to give slightly small quadrupole 
deformation. 

The final continuum state $\Psi_f$ regular at the origin 
is obtained as follows. 
In a single-channel approximation with the adiabatic
potential~(\ref{ave.pot}), the equation of motion for 
$\psi(E,\rho)=\rho^{5/2}\Psi_f$ with energy $E$ is derived 
using Eq.~(\ref{kinetic.op}) as  
\begin{align}
\left(\frac{d^2}{d\rho^2}-\frac{\Lambda(\Lambda+1)}{\rho^2}+k^2-
\frac{2m_{\alpha}}{\hbar^2}V(\rho)\right) \psi(E,\rho)=0,
\label{radial.eq}
\end{align}
where $k^2=2m_{\alpha}E/\hbar^2$ and $\Lambda=3/2$. To obtain 
a solution $\psi(E,\rho)$, we note that  
$V(\rho)$ approaches $Z_{\rm eff}e^2/\rho$ 
($Z_{\rm eff}=12$) 
for large $\rho$. Therefore the solution of Eq.~(\ref{radial.eq}) at 
large $\rho$ can be expressed as a combination of regular and 
irregular Coulomb wave functions $F_{\Lambda}(\eta, \rho)$ and 
$G_{\Lambda}(\eta, \rho)$, where $\eta=m_{\alpha}Z_{\rm eff}e^2/\hbar^2 k$. Note, however, that 
$\psi(E,\rho)$ in the asymptotic region is subject to the $\delta(E-E')$ normalization of $\Psi_f$.
That is, $\psi(E,\rho)$ satisfies  
\begin{align}
\psi(E,\rho) \to \sqrt{\frac{2m_{\alpha}}{\pi^4 \hbar^2 k}}
\left[ \cos \delta F_{\Lambda}(\eta, \rho) +
\sin \delta G_{\Lambda}(\eta, \rho)\right]
\end{align}
for large $\rho$. Here $\delta$ is the three-body phase shift. 

\subsection{$E2$ strength function and reaction rate}

\begin{figure}[b]
\begin{center}
\epsfig{file=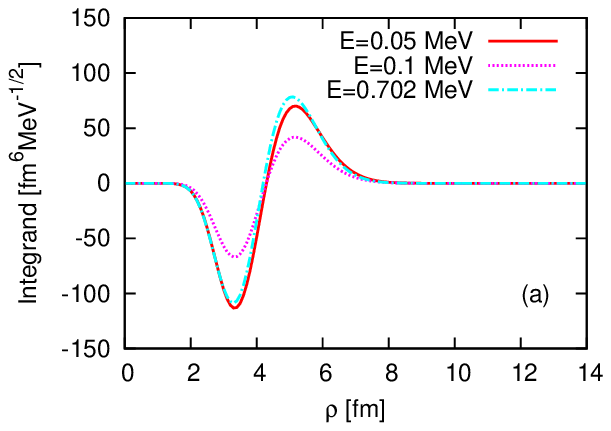,scale=1.3}
\epsfig{file=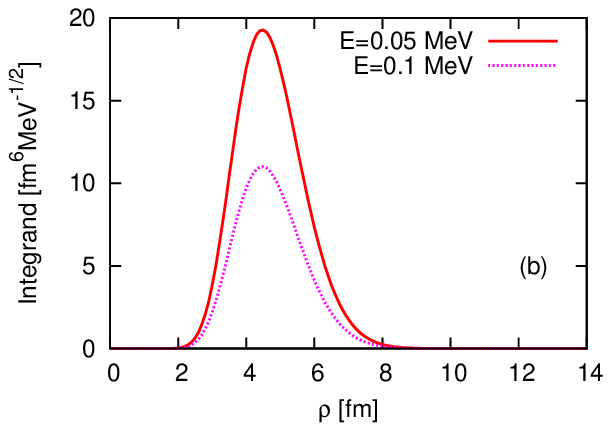,scale=1.3}
\caption{
(Color online) The integrand of 
$I_{i \to f}(E)$ for AB(A$^{\prime}$) (a) and AB(D) (b) potentials 
as a function of $\rho$. Plotted curves are magnified by 
$3\cdot 10^{19}$ for $E=0.05$\, MeV and by $3\cdot 10^{12}$ for $E=0.1$\, MeV, respectively. No multiplication is 
made for $E=0.702$\,MeV curve.  }
\label{integrand.ab}
\end{center}
\end{figure}

The $E2$ strength function~(\ref{strength.fn}) for the transition from the $2^+$ 
state to the continuum is given by 
\begin{align}
S_{E2}(E_{\gamma})=\frac{5\pi^2e^2}{96}a^5 [I_{i \to f}(E)]^2
\label{SE2.gamma}
\end{align}
with the radial integral between the initial and final states 
\begin{align}
I_{i \to f}(E)=\int_0^{\infty}d\rho \, \rho^{13/2}\psi(E,\rho)\exp(-{\textstyle{\frac{1}{2}a\rho^2}}).
\label{integrand}
\end{align}
The integral $I_{i \to f}(E)$ plays a decisive role  
to determine the reaction rate. 
The amplitude of $\psi(E,\rho)$ in the region that contributes  
to the integral is determined by the potential 
$V(\rho)$ from the asymptotic region down to the internal 
region as the normalization of $\psi(E,\rho)$ is fixed  asymptotically.
The lower the potential barrier, the larger the 
reaction rate.

Panels (a) and (b) of Fig.~\ref{integrand.ab} display the integrand of $I_{i \to f}(E)$ for AB(A$^{\prime}$) and AB(D) potentials, respectively. The shape of the integrand looks very similar in each case. 
It is remarkable  
that the integrand of AB(A$^{\prime}$) has a node that is almost energy-independent,  while the one of AB(D) 
has no such node. The reason is that the adiabatic potential 
calculated with AB(A$^{\prime}$) is deep enough to 
accommodate a $0^+$ bound state, so that the continuum 
wave function has to be orthogonal to that bound state. 
Since the AB(D) potential supports no bound state, however, the continuum wave function in that adiabatic potential can reach the inner region without the orthogonality constraint. Moreover, the magnitude of 
the integrand is quite different. With the AB(A$^{\prime}$) interaction, 
we expect a strong cancellation for $I_{i \to f}(E)$, whereas no such cancellation occurs in AB(D). The  
magnitude of the integrand of AB(A$^{\prime}$) is much 
larger than that of AB(D).

\begin{figure}
\begin{center}
\epsfig{file=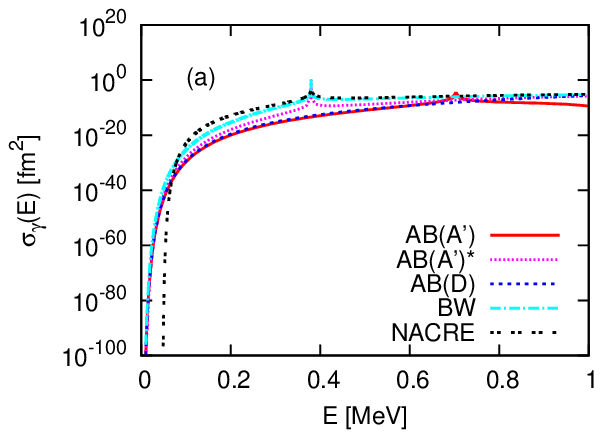,scale=1.4}
\epsfig{file=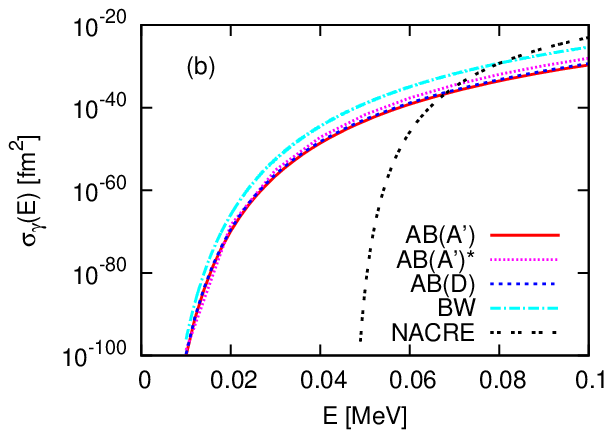,scale=1.4}
\caption{
(Color online) Photoabsorption cross section of the $2^+$ state of $^{12}$C 
into 3$\alpha$ $0^+$ continuum state with the $3\alpha$ 
energy $E$. Panel (b) enlarges panel (a) at low energies. The photoabsorption cross sections calculated  according to NACRE model~\cite{nacre} and BW formula (Eq.~(4) of Ref.~\cite{garrido11}) for the direct process are also shown.  }
\label{sigma.gamma}
\end{center}
\end{figure}

Figure~\ref{sigma.gamma} displays the photoabsorption 
cross section $\sigma_{\gamma}(E)$ calculated with the
AB(A$^{\prime}$) and AB(D)  
potentials together with NACRE and BW cross sections. We also show   
AB(A$^{\prime}$)$^{\ast}$ calculation that uses the AB(A$^{\prime}$) 
potential with the strength of the three-body force being adjusted to reproduce the Hoyle resonance energy. The calculated resonance width is, however, too small to be 
compared to experiment. 
Table~\ref{sigma-rate.comp} compares the low-energy $\sigma_{\gamma}$ 
values calculated by the several models in ratio to NACRE cross sections~\cite{garrido11}.  The present result is between BW and Faddeev. In the BW model, the low-energy $\sigma_{\gamma}$ is controlled by the width 
$\Gamma_{3\alpha}$, which is not yet determined experimentally and 
is simply assumed to be equal to the total width $\Gamma$ of the Hoyle
state. However, since the 
Hoyle state decays predominantly via the $\alpha$+$^8$Be$(0^+)$ channel, 
$\Gamma_{3\alpha}$ is probably considerably smaller than 
$\Gamma$~\cite{kirsebom12}. 
Thus the BW value may be considered as the upper limit of the 
direct process at very low energies. 
As shown in the table, all of $\sigma_{\gamma}$ values calculated by different models are 
very much enhanced compared to NACRE at  
0.05\,MeV, but there is a big difference among them: 
The HHR result is much larger than the BW value, 
while the Faddeev gives smaller value. At $E=0.1$\,MeV the 
difference among the models becomes smaller.

\begin{table*}
\begin{center}
\caption{Comparison of the photoabsorption 
cross section $\sigma_{\gamma}$ and the Maxwell-Boltzmann  energy-averaged triple-alpha reaction rate 
$N_A^2\langle R_{\alpha \alpha \alpha} \rangle$.  
AB(A$^{\prime}$) potential is used except for BW and CDCC.  Values in
 parentheses are obtained with AB(D) potential, while those with
 $^{\ast}$ are obtained with 
AB(A$^{\prime}$)$^{\ast}$ potential. The cross sections are 
given in ratio to NACRE~\cite{nacre}, which 
is 1.6$\times 10^{-84}$ at $E=0.05$ 
and 9.7$\times 10^{-24}$\,fm$^2$ at $E=0.1$\,MeV, respectively. The reaction rate is also   
given in ratio to NACRE, which   
is 2.9$\times 10^{-71}$ at $T=0.01$ 
and 1.5$\times 10^{-47}$\,cm$^6$s$^{-1}$mol$^{-2}$ at $T=0.03$\,GK, respectively. }
\label{sigma-rate.comp}
\begin{tabular}{cccccccc}
\hline\hline
$\sigma_{\gamma}$&&&&&&& \\
\hline
&$E$(MeV)&&Present &BW~\cite{garrido11}&HHR~\cite{nguyen12}&Faddeev~\cite{ishikawa13}&\\
\hline
&0.05 && $4\cdot 10^{40}$($9\cdot 10^{40}$) & $6\cdot 10^{44}$ & $2\cdot 10^{52}$ & $1\cdot 10^{38}$($8\cdot 10^{37}$)& \\ 
& && ${1\cdot 10^{42}}^{\ast}$ &  &  & & \\ 
&0.1 && $2\cdot 10^{-7}$($5\cdot 10^{-7}$) & $5\cdot 10^{-3}$ & $4\cdot 10^{-3}$ & $3\cdot 10^{-8}$($3\cdot 10^{-8}$)& \\
& && ${1\cdot 10^{-5}}^{\ast}$ & & & & \\
\hline\hline
$N_A^2\langle R_{\alpha \alpha \alpha} \rangle$&&&&&&&\\
\hline
&$T$(GK)&&Present &BW~\cite{garrido11}&HHR~\cite{nguyen12}&Faddeev~\cite{ishikawa13}&CDCC~\cite{ogata09}\\
\hline
&0.01 && $2\cdot 10^{3}$($4\cdot 10^{3}$) & $3\cdot 10^{7}$ & $3\cdot 10^{18}$ & $1\cdot 10^{1}(5\cdot 10^0)$ & $4\cdot 10^{26}$\\ 
& && ${2\cdot 10^{3}}^{\ast}$ &  &  & & \\ 
&0.03 && $2\cdot 10^{0}$($5\cdot 10^{0}$) & $5\cdot 10^{4}$ & $1\cdot 10^{7}$ & $2\cdot 10^{0}(9\cdot 10^{-1})$ & $2\cdot 10^{18}$ \\
& && ${2\cdot 10^{0}}^{\ast}$ &  &  & & \\ 
\hline\hline
\end{tabular}
\end{center}
\end{table*}

\begin{figure}
\begin{center}
\epsfig{file=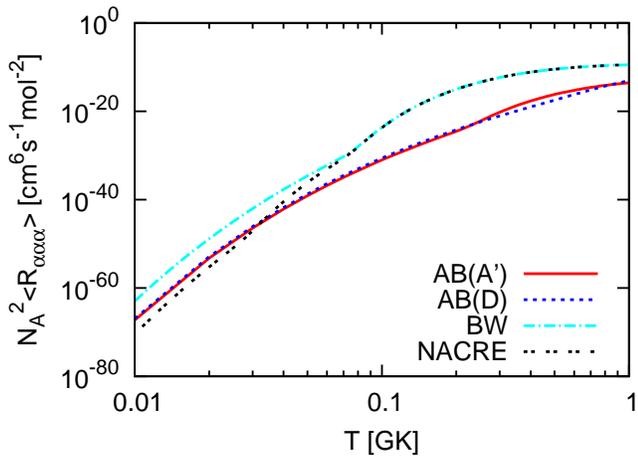,scale=1.4}
\caption{
(Color online) The same as Fig.~\ref{sigma.gamma} but for the Maxwell-Boltzmann  energy-averaged triple-$\alpha$ reaction rate as a function of temperature $T$.}
\label{triplerate}
\end{center}
\end{figure}

Figure~\ref{triplerate} compares  
the energy averaged triple-$\alpha$ reaction rate with NACRE. 
Both AB(A$^{\prime}$) and AB(A$^{\prime}$)$^{\ast}$ give almost 
the same reaction rate. Because the properties of the Hoyle resonance
are not reproduced in
the present calculation, the reaction rate above $T=0.1$\,GK is much
smaller than the NACRE rate. 
Table~\ref{sigma-rate.comp} compares 
our reaction rates at 0.01 and 0.03\,GK with those obtained by BW, HHR, Faddeev, and CDCC calculations. The huge enhancement 
of CDCC calculation is not supported by any other calculations. However, the enhancement compared to NACRE is 
still at variance depending on the model. 
We see that both present and Faddeev 
calculations give rather close results at the two 
temperatures. Compared to these, HHR rate is larger 
by 10$^{15}$ and BW rate is larger by 
10$^{4}$ at 0.01\,GK. Though the difference tends to decrease at 0.03\,GK, there is still a
difference by about 10$^7$ at maximum.

\subsection{Symmetrization effects}
\begin{figure}
\begin{center}
\epsfig{file=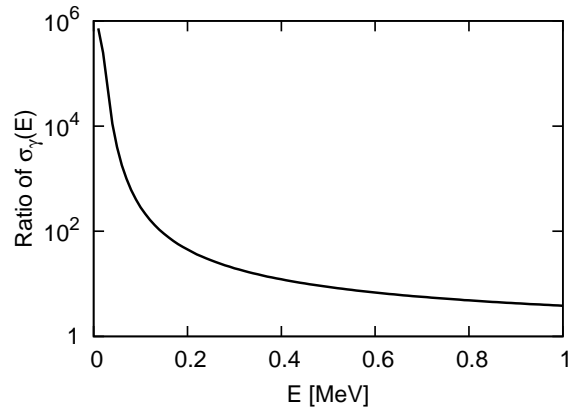,scale=1.25}
\caption{
Ratio of the photoabsorption cross sections calculated with the non-symmetrized and symmetrized wave functions as a function of the $3\alpha$ energy $E$. The adiabatic potential is obtained with 
the $\alpha \alpha$ potential of Ref.~\cite{ogata09}. }
\label{sigmaratio.ogata}
\end{center}
\end{figure}

\begin{figure}
\begin{center}
\epsfig{file=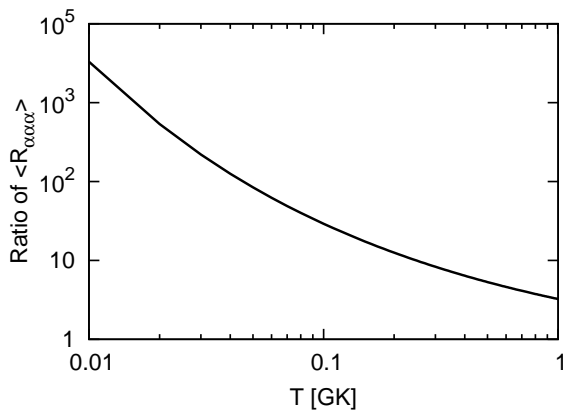,scale=1.25}
\caption{
The same as 
Fig.~\ref{sigmaratio.ogata} but for the ratio of 
the Maxwell-Boltzmann  energy-averaged triple-$\alpha$ reaction rate as a function of temperature $T$. }
\label{R-tempratio.ogata}
\end{center}
\end{figure}

We examine the extent to which the neglect of symmetrization changes the reaction rate. 
We use the $\alpha \alpha$ potential of Ref.~\cite{ogata09} and 
calculate the adiabatic 
potential using the wave function of Eq.~(\ref{gcmwf}) 
itself (non-symmetrized version) and its symmetrized 
wave function (symmetrized version).  
Figure~\ref{sigmaratio.ogata} compares the 
photoabsorption cross sections obtained with the 
adiabatic potential of the non-symmetrized version with the one of the symmetrized version. 
The non-symmetrized cross section is more than 
10$^6$ times larger than the symmetrized one at, e.g., 0.01\,MeV, but the enhancement of the energy-averaged 
reaction rate 
$\langle R_{\alpha \alpha \alpha} \rangle$ is more moderate as shown in 
Fig.~\ref{R-tempratio.ogata}. At very low temperature where 
the non-resonant contribution is expected to be important, 
the enhancement is on the order of 10$^3$ at $T=0.01$\,GK. A part of the reason for the huge enhancement reported in 
Ref.~\cite{ogata09} is due to this neglect of the symmetrization, but its effect is not 
large enough to account for such huge enhancement as  $10^{26}$ at 0.01\,GK.

\subsection{Discussion of the literature}
\label{discuss.lit}
In the following we attempt at understanding possible 
reasons for the large photoabsorption 
cross sections of CDCC and HHR. 
The calculations of Refs.~\cite{ogata09,ishikawa13} are 
performed in the Jacobi coordinates, $\bm r=\sqrt{2}\bm x_1$ and $\bm R=\sqrt{3/2}\bm x_2$. With these coordinates the Coulomb potential~(\ref{coulomb.3alpha}) is expanded in multipoles as   
\begin{align}
V_C=
\frac{4e^2}{r}+\frac{8e^2} {R_>}\sum_{\ell={\rm even}}\left(\frac{R_<}{R_>}\right)^{\ell} P_{\ell}(\hat{\bm r}\cdot \widehat {\bm R}),
\label{8Be-alpha}
\end{align}
where $R_> (R_<)$ denotes the larger (smaller) between $R$ and $r/2$, and $P_{\ell}$ is the 
Legendre polynomial of degree ${\ell}$. Thus the coupling between $\bm r$ and $\bm R$ is 
always present everywhere, and it is crucial to take care of 
such couplings in the calculation.
The 3\,$\alpha$ continuum wave function may be written in the
spirit of CDCC as 
\begin{align}
\Psi \sim \sum_i \frac{u_i(r)}{r}\frac{\chi_i(R)}{R}
[Y_{\ell_i}(\hat{\bm r}) Y_{\ell_i}(\hat{\bm R})]_{00}+\Psi_{\rm dc},
\label{cdcc.general}
\end{align}
where $u_i(r)Y_{\ell_i}(\hat{\bm r})$ is the $\alpha \alpha$ 
continuum-discretized  
state of $i$th bin and $\Psi_{\rm dc}$, vanishing asymptotically, 
stands for square-integrable distorted
components other than the first term. 
Neither high-partial waves nor $\Psi_{\rm dc}$ is included 
in Ref.~\cite{ogata09}. 

The contribution of the Coulomb potential to the coupling potential reads 
\begin{align}
V_{ij}^C(R)&=\frac{8e^2}{R}\sum_{\ell={\rm even}}C^{\ell}_{ij} 
\Bigg\{
\int_0^{2R}dr u_i^*(r)u_j(r)\big({\textstyle{\frac{r}{2R}}}\big)^{\ell}
\notag \\
&\ \ +\int_{2R}^{\infty}dr u_i^*(r)u_j(r)
\big({\textstyle{\frac{2R}{r}}}\big)^{\ell+1}\Bigg\},
\end{align}
where $C_{ij}^{\ell}$ is a matrix element of type~(\ref{yyp}).
The coupling potential in general 
never vanishes even for large $R$, but  
if only $S$ wave is included, 
no coupling arises at large $R$. The truncation to $S$ wave only  
will thus lead to enhancing the reaction rate to some extent. At least 
$D$ wave has to be included. 

More important is the role of $\Psi_{\rm dc}$. There are a number
of cases that demonstrate the importance of $\Psi_{\rm dc}$ 
to obtain converged solutions for scattering and radiative capture reactions 
(see, e.g., Refs.~\cite{aoyama12,quaglioni08,arai11,neff11}). The form of 
$\Psi_{\rm dc}$ depends on the problem concerned. In the 
present case a primary concern is to take proper account of the Coulomb 
potential of 3\,$\alpha$ particles.  
Most of $u_i(r)$'s included in the CDCC 
calculation~\cite{ogata09} are spread to 
large distances, so that the first term of Eq.~(\ref{cdcc.general})
alone 
may not be flexible enough to represent the damping of the amplitude of 
$\Psi$ in the region where the photoabsorption occurs. 
If this is the case, an explicit 
inclusion of some distorted configurations is needed to 
make $\Psi$ realistic, which would lead to a smaller reaction rate.


In the HHR calculation~\cite{nguyen12} the hyperspherical 
coordinate is used instead of  
$\bm r$ and $\bm R$. Choosing the 
$R$-matrix radius to be $\rho=25$\,fm, the 3\,$\alpha$ 
wave function inside the region is expanded  
as a superposition of the HH functions with 
$K_{\rm max}\approx 26$, and then it is propagated to the 
asymptotic region. The $E2$ strength function of HHR is much larger than the 
present one at $E\leq 0.05$\,MeV. This indicates that the 
amplitude of the HHR continuum state is very much enhanced 
at the low energies. 
One  possible 
reason for this may be in the tail behavior of the Coulomb potential.  
In the HHR the Coulomb coupling is taken into account up to 
$\rho=400$\,fm keeping $K_{\rm max}\approx 26$ and, after that the off-diagonal coupling is screened up to 1500\,fm. This procedure together with the $R$-matrix propagation of $K_{\rm max}\approx 26$ truncation might lead to the 
tail behavior that is different from ours.  
As shown in Table~\ref{HHexpansion} and Fig.~\ref{kdistr} 
of Appendix~\ref{hhe}, the convergence of the HH expansion becomes 
slower as the size of 3\,$\alpha$ system becomes larger. This suggests 
that the
$K$ truncation made in the innermost region 
may result in preventing the continuum wave function 
from spreading to many more $K$ components during the process of 
$R$-matrix propagation.

\section{Conclusion}
\label{conclusion}

We address the triple-alpha reaction at very low energies to obtain 
its reaction rate 
below 0.1\,GK. On the basis of the direct capture  
process of three $\alpha$ particles, we discuss the potential barrier through which 
3\,$\alpha$ particles penetrate and fuse to make the radiative transition to the 2$^+$ 
state of $^{12}$C. The general properties of the 3\,$\alpha$ Coulomb potential 
that dominates the barrier at large distances are carefully examined in hyperspherical coordinates 
and the minimum Coulomb barrier is established. 
Since the hyperspherical harmonics (HH) expansion is slow as the size of 3\,$\alpha$ 
system expands, the adiabatic potential barrier as a function of 
the hyperradius is estimated by averaging the 
potential energy expectation values of various geometric configurations specified 
by Gauss wave packets. 

Our results on the triple-alpha rate do not support the large continuum-discretized coupled-channels (CDCC) values at 
0.01\,GK, but fall between those of the Breit-Wigner model and the Faddeev method. 
We attempt at understanding 
possible mechanism of how the large rate is obtained in the CDCC  
and HH basis $R$-matrix calculations. 
Though it is simply assumed as the average potential 
energy in the present study, the adiabatic potential barrier should in principle be obtained 
by taking into account the coupling of 
various configurations. A study along this extension will be interesting. 
A final goal will be a microscopic study of the triple-alpha reaction process.

\section*{Acknowledgments}
One of the authors (Y.S.) thanks E. Garrido for several illuminating correspondences as well as for his codes of the Breit-Wigner cross sections. He is 
indebted to D. Baye for useful discussions and the support for his stay
at ULB, June 2013, where a part of this work was 
performed. The work of Y.S. is supported in part by   
Grant-in-Aid for Scientific Research (No. 24540261) of the Japan Society for the Promotion of Science. 
P.D. is Directeur de Recherches F.R.S.-FNRS.

\appendix

\section{Symmetric hyperspherical harmonics}
\label{app.symm.pol}
 
An orthonormal set in the HH is constructed from the eigenfunction 
of the hypermomentum operator ${\cal K}^2$ of Eq.~(\ref{hyp.momentum}). 
The normalized eigenfunction with 
the eigenvalue $K(K+4)$ is given by 
\begin{align}
{F}_{KLM}^{\ell_1 \ell_2}(\Omega_x)=\phi_{K}^{\ell_1
 \ell_2}(\alpha)[Y_{\ell_1}(\hat{\bm x}_1)Y_{\ell_2}(\hat{\bm
 x}_2)]_{LM},
\label{eigfn.K}
\end{align}
where $K$ is an integer called the hypermomentum and 
$\phi_{K}^{\ell_1 \ell_2}(\alpha)$ is an orthonormal 
function 
\begin{align}
&\phi_{K}^{\ell_1 \ell_2}(\alpha)\notag \\
&={\cal N}_K^{\ell_1 \ell_2}
\cos^{\ell_1}\alpha  \sin^{\ell_2}\alpha 
G_n(\ell_1+\ell_2+2,\ell_2+\textstyle{\frac{3}{2}}; \sin^2\alpha) 
\end{align}
with the normalization constant 
\begin{equation}
{\cal N}_K^{\ell_1 \ell_2}=\sqrt{\frac{2(K+2)\Gamma(\ell_1+\ell_2+n+2)\Gamma(\ell_2+n+
\textstyle{\frac{3}{2}})}{n!\Gamma(\ell_1+n+\textstyle{\frac{3}{2}})
[\Gamma(\ell_2+\textstyle{\frac{3}{2}})]^2}},
\end{equation}
where $n$ is an integer given by $n$=$(K-\ell_1-\ell_2)/2$, and $G_n$ is the Jacobi polynomial that is expressed in terms of the Gauss hypergeometric series as follows
\begin{align}
&G_n(\ell_1+\ell_2+2,\ell_2+{\textstyle{\frac{3}{2}}};z^2)\notag \\
&=F(-n,\ell_1+\ell_2+n+2,\ell_2+{\textstyle{\frac{3}{2}}};z^2).
\end{align}
Note that for $L=0$, $\ell_1$ and $\ell_2$ are equal and the allowed values of $K$ are even.  

One can define angles $\Omega_y$ in $\bm y$ coordinate in completely 
the same way as $\Omega_x$.  The function 
${F}_{KLM}^{\ell_1 \ell_2}(\Omega_x)$, when expressed in terms of the coordinate $\Omega_y$, becomes a linear 
combination of functions ${F}_{KLM}^{\ell_1' \ell_2'}(\Omega_y)$ 
with $KLM$ being unchanged. The expansion coefficient 
is Raynal-Revai coefficient 
$\langle {F}_{KLM}^{\ell_1' \ell_2'}(\Omega_y)|
{F}_{KLM}^{\ell_1 \ell_2}(\Omega_x)\rangle$~\cite{raynal70}. 

Let us focus on a system of three identical particles. We want to construct a symmetric function for given $K$ and $L$ values by 
\begin{align}
{F}_{KLM}^{\gamma}=\sum_{\ell_1 \ell_2 }C_{\ell_1 \ell_2}^{\gamma} 
{F}^{\ell_1 \ell_2}_{KLM}(\Omega_x),
\label{sym.basis}
\end{align}
where $\gamma$ is a label to distinguish different 
symmetric functions. 
The coefficient $C_{\ell_1 \ell_2}^{\gamma}$ is determined by solving the linear equation 
\begin{align}
&\langle {F}^{\ell_1 \ell_2}_{KLM}(\Omega_y)|{ F}_{KLM}^{\gamma}\rangle
\notag \\
&=\sum_{\ell_1' \ell_2'}
 \langle {F}^{\ell_1 \ell_2}_{KLM}(\Omega_y)| 
{F}^{\ell_1' \ell_2'}_{KLM}(\Omega_x)\rangle C_{\ell_1' \ell_2'}^{\gamma}\notag \\
&=C_{\ell_1 \ell_2}^{\gamma},
\end{align} 
which must be satisfied for all possible values of $\ell_1$ and $\ell_2$ compatible with $K$ and $L$. 
Here the last equality is due to the fact that $\Omega_x$ in 
Eq.~(\ref{sym.basis}) may be replaced with $\Omega_y$ because 
${F}_{KLM}^{\gamma}$ is a symmetric function. 

We have constructed the symmetric basis functions for $L=0$. 
The number of symmetric functions is given by 
$[K/12]+1$. 
When $K$ is  
a multiple of 12 plus 2, however, the number 
is $[K/12]$. Here $[c]$ is the Gauss symbol,  
indicating the greatest integer that does not exceed $c$. 
We have no symmetric HH function for $K=2$.

\section{Matrix element of Coulomb potential in HH basis}
\label{coulomb.m.e}

In this appendix we generalize the masses and charges of three particles. 
Let $A_1, A_2, A_3$ be the mass ratios $A_i=m_i/m$ of the 
three particles, where $m$ is some unit mass, and 
$Z_1e, Z_2e, Z_3e$ be the charges of the three particles. 
Let us define the coordinates $\bm x_1$ and $\bm x_2$ by 
\begin{align}
&\bm x_1=\sqrt{A_{1,2}}(\bm r_1-\bm r_2),\notag \\
&\bm x_2=\sqrt{A_{12,3}}\left(\frac{A_1\bm r_1+A_2\bm r_2}{A_1+A_2}-\bm r_3\right),
\end{align}
where $A_{i,j}=A_iA_j/(A_i+A_j), A_{ij,k}=(A_i+A_j)A_k/(A_i+A_j+A_k)$. In Eq.~(\ref{jacobi.coord}) $m$ is 
taken to be $m_{\alpha}$. The hyperradius $\rho$ and 
hyperangle $\alpha$ are defined as before by $x_1=\rho \cos \alpha$, 
$x_2=\rho \sin \alpha$. 
 
The Coulomb potential $V_{C}$ acting among the particles 
is expressed in terms of $\rho$ and $\Omega_x$ as follows:  
\begin{align}
V_{C}&=\frac{Z_1Z_2e^2}{|{\bm r}_1-{\bm r}_2|}
+\frac{Z_2Z_3e^2}{|{\bm r}_2-{\bm r}_3|}
+\frac{Z_3Z_1e^2}{|{\bm r}_3-{\bm r}_1|}\notag \\
&=\frac{e^2}{\rho}Q(\Omega_x)
\end{align}
with the charge factor  operator 
\begin{align}
Q(\Omega_x)&=\frac{Z_1Z_2\sqrt{A_{1,2}}}{\cos \alpha}\notag \\
&+\frac{Z_2Z_3}
{|\frac{\sqrt{A_{1,2}}}{A_2}\cos \alpha \, \hat{\bm x}_1-
\frac{1}{\sqrt{A_{12,3}}}\sin \alpha \, \hat{\bm x}_2|}
\notag \\
&+\frac{Z_3Z_1}
{|\frac{\sqrt{A_{1,2}}}{A_1}\cos \alpha \, \hat{\bm x}_1+
\frac{1}{\sqrt{A_{12,3}}}\sin \alpha \, \hat{\bm x}_2|}.
\label{chargefactor}
\end{align}

We calculate the matrix element of Eq.~(\ref{chargefactor}) 
in the HH functions.  
The second term of $Q(\Omega_x)$ is expanded as 
\begin{align}
&\frac{Z_2Z_3}
{|\frac{\sqrt{A_{1,2}}}{A_2}\cos \alpha \, \hat{\bm x}_1-
\frac{1}{\sqrt{A_{12,3}}}\sin \alpha \, \hat{\bm x}_2|}\notag \\
&=
Z_2Z_3\sum_{\ell}\frac{{s_<}^{\ell}}{{s_>}^{\ell+1}}P_{\ell}(\cos
\omega),
\end{align}
where $s_< \, (s_>)$ denotes the smaller (larger) one of 
$\frac{\sqrt{A_{1,2}}}{A_2}\cos \alpha$ and
$\frac{1}{\sqrt{A_{12,3}}}\sin \alpha$, and $\omega$ is the angle
between $\hat{\bm x}_1$ and $\hat{\bm x}_2$.
The matrix element of $P_{\ell}(\cos \omega)$ is 
\begin{align}
& \langle [Y_{\ell_1}(\hat{\bm x}_1)Y_{\ell_2}(\hat{\bm
 x}_2)]_{LM}|P_{\ell}(\cos \omega)| [Y_{\ell_1'}(\hat{\bm x}_1)Y_{\ell_2'}(\hat{\bm
 x}_2)]_{LM}\rangle
\notag \\
& = 4\pi \frac{(-1)^{\ell}}{2\ell +1}\sqrt{\frac{2\ell_1+1}{2\ell_1'+1}}U(\ell_1 \ell L
 \ell_2'; \ell_1' \ell_2)\notag \\
&\ \ \times C(\ell \ell_1'; \ell_1)C(\ell \ell_2'; \ell_2)
\label{yyp}
\end{align}
with 
\begin{equation}
C(\ell_1 \ell_2;\ell_3)=\sqrt{\frac{(2\ell_1+1)(2\ell_2+1)}{4\pi (2\ell_3+1)}}
\langle \ell_10\ell_20\vert \ell_30 \rangle,
\end{equation}
where $U$ is a Racah or 6$j$ coefficient in a unitary form. 
Similarly the third term of $Q(\Omega_x)$ is expanded as 
\begin{align}
&\frac{Z_3Z_1}
{|\frac{\sqrt{A_{1,2}}}{A_1}\cos \alpha \, \hat{\bm x}_1+
\frac{1}{\sqrt{A_{12,3}}}\sin \alpha \, \hat{\bm x}_2|}\notag \\
&=Z_3Z_1\sum_{\ell}\frac{{t_<}^{\ell}}{{t_>}^{\ell+1}}P_{\ell}(-\cos \omega),
\end{align}
where $t_< \, (t_>)$ denotes the smaller (larger) one of 
$\frac{\sqrt{A_{1,2}}}{A_1}\cos \alpha$ and
$\frac{1}{\sqrt{A_{12,3}}}\sin \alpha$. 

Combining the above results leads to the matrix element $Q_{cc'}$ ($c$=$(K \ell_1 \ell_2)$):
\begin{eqnarray}
Q_{cc'}=\langle \phi_K^{\ell_1 \ell_2}(\alpha)|
Z_L^{\ell_1 \ell_2, \ell_1' \ell_2'}(\alpha)|  \phi_{K'}^{\ell_1'
\ell_2'}(\alpha)\rangle,
\end{eqnarray}
where $\langle f(\alpha)|g(\alpha)\rangle =\int_0^{\pi/2} d\alpha \cos^2\alpha \sin^2\alpha 
f(\alpha)g(\alpha)$ and  
\begin{align}
&Z_L^{\ell_1 \ell_2, \ell_1' \ell_2'}(\alpha) \notag \\
&=Z_1Z_2\sqrt{A_{1,2}}
\delta_{\ell_1,\ell_1'}\delta_{\ell_2,\ell_2'}\frac{1}{\cos \alpha}
\notag \\
&+\sum_{\ell}\langle [Y_{\ell_1}(\hat{\bm x}_1)Y_{\ell_2}(\hat{\bm
 x}_2)]_{LM}|P_{\ell}(\cos \omega)| [Y_{\ell_1'}(\hat{\bm x}_1)Y_{\ell_2'}(\hat{\bm
 x}_2)]_{LM}\rangle
\notag \\
& \ \ \times 
\left[Z_2Z_3\frac{{s_<}^{\ell}}{{s_>}^{\ell+1}}
+(-1)^{\ell}Z_3Z_1\frac{{t_<}^{\ell}}{{t_>}^{\ell+1}}
\right].
\end{align}
The range of $\ell$ in the above sum is limited by the triangular condition of 
$\ell_1, \ell_1', \ell$ and $\ell_2, \ell_2', \ell$. Also 
both $\ell_1+\ell_1'+\ell$ and $\ell_2+\ell_2'+\ell$ must be even.

In case of three $\alpha$ particles, $s$ and $t$ are given as 
\begin{align}
&s_<=t_<=\Bigg\{
\begin{array}{ll}
\frac{\sqrt{3}}{2\sqrt{2}}\sin \alpha & \ 0\leq \alpha \leq \frac{\pi}{6} \\
\frac{1}{2\sqrt{2}}\cos \alpha & \ \frac{\pi}{6}\leq \alpha \leq \frac{\pi}{2} \\
\end{array}, \notag \\
&s_>=t_>=\Bigg\{
\begin{array}{ll}
\frac{1}{2\sqrt{2}}\cos \alpha & \ 0\leq \alpha \leq \frac{\pi}{6} \\
\frac{\sqrt{3}}{2\sqrt{2}}\sin \alpha & \ \frac{\pi}{6}\leq \alpha \leq \frac{\pi}{2} \\
\end{array}.
\end{align}

A more elegant way to calculate the matrix elements of the second and third terms in 
Eq.~(\ref{chargefactor}) is to
transform ${\cal F}_{KLM}^{\ell_1 \ell_2}$ in  $\bm x$ coordinate 
to those of  $\bm y$ and $\bm z$
coordinates using the Raynal-Revai coefficients. Then we need to 
consider the very simple matrix element of type of the first term only.

\section{Convergence of hyperspherical harmonics expansion}
\label{hhe}

A three-body problem is often solved in the HH method. The accuracy of
such solution depends on whether or not the HH
functions with sufficiently large $K$ values are included in the 
calculation. We here examine the convergence of the HH expansion.  
To give specific examples, we take the same wave function 
as Eq.~(\ref{gcmwf}) with a slight modification of angular momentum
projection.  Projecting $S$-waves 
for both $\bm x_1$ and $\bm x_2$ coordinates, we have the 
following shifted Gauss function 
\begin{align}
&\Phi_{00}(s_1,s_2,\bm x)\notag \\
&={\cal N}(s_1,s_2)
\exp(-{\textstyle{\frac{1}{2}\beta(x_1^2+s_1^2)}})i_0(\beta s_1x_1)
 \notag \\
& \times
 \exp(-{\textstyle{\frac{1}{2}\beta(x_2^2+s_2^2)}})i_0(\beta s_2x_2)
[Y_0(\hat{\bm x}_1)Y_0(\hat{\bm x}_2)]_{00},
\label{shifted.gauss}
\end{align}
where $i_0(x)=\sinh x /x$ and ${\cal N}(s_1,s_2)$ is the normalization constant
\begin{align}
&[{\cal N}(s_1,s_2)]^{-1}\notag \\
&=\frac{\sqrt{\pi}}{4\beta^{5/2} s_1s_2}
\sqrt{{\textstyle{(1-\exp(-\beta s_1^2))(1-\exp(-\beta s_2^2))}}}. 
\end{align}
The function~(\ref{shifted.gauss}) has 
a peak at $(x_1, x_2) \sim (s_1, s_2)$, so that 
we obtain various configurations of 3\,$\alpha$ particles by changing 
$s_1,\, s_2$. 

Expanding $\Phi_{00}(s_1,s_2,\bm x)$ in the HH function as 
\begin{align}
\Phi_{00}(s_1,s_2,\bm x)=\sum_{K} f_K(s_1,s_2,\rho){ F}^{00}_{K00}(\Omega_x)
\end{align}
with 
\begin{align}
f_K(s_1,s_2,\rho)=\langle {F}^{00}_{K00}(\Omega_x)|
 \Phi_{00}(s_1,s_2,\bm x)\rangle,
\end{align}
we obtain the probability of finding $K$ component in 
$\Phi_{00}(s_1,s_2,\bm x)$ by a squared norm
\begin{align}
 ||f_K(s_1,s_2)||^2=\int_0^{\infty}d\rho \, \rho^5
[f_K(s_1,s_2,\rho)]^2,
\end{align}
which satisfies 
$\sum_K ||f_K(s_1,s_2) ||^2=1$. The distribution of  
$||f_K(s_1,s_2) ||^2$ with respect to $K$ serves a measure 
of convergence of the HH expansion. Note that
$||f_K(s_1,s_2)||^2=||f_K(s_2,s_1)||^2$ for the wave 
function~(\ref{shifted.gauss}).

\begin{table*}
\begin{center}
\caption{Probability in \% of finding the hyperspherical harmonics with
 hypermomentum $K$ in the shifted Gaussian~(\ref{shifted.gauss}).  The 
 probability that does not exceed 0.1 \% is indicated by the line --.}
\label{HHexpansion}
\begin{tabular}{cccccccccccccccccccc}
\hline\hline
$\sqrt{\gamma_1^2+\gamma_2^2}$& $(\gamma_1,\, \gamma_2)$ && & & & & & & & & $K$ & & & & & & & & \\
\cline{3-20}
 &&& 0& 2& 4& 6& 8& 10& 12& 14& 16 & 18 & 20 &22&24&26&28&30 &\, $\geq 32$\\
\hline
4.95&(3.5,\, 3.5) && 77.7 & -- & 20.6 & -- & 1.6 & -- & -- & -- & --& -- &
						 --&--&--&--&--&--&\, -- \\ 
21.3&(14,\, 16) && 20.7 & 1.4 &16.6 &4.8 & 10.5 &7.9 & 4.9 &9.2&1.4&8.3&--
				&5.9 &0.1 &3.4 &0.5&1.6 &\, 2.7  \\ 
21.3&(3.5,\, 21) && 2.2 & 7.7 &13.7 &17.3 & 16.9 &13.0 & 7.5 &2.8&0.3&0.2&1.6
				&3.2 &4.0 &3.7 &2.7&1.6 &\, 1.6  \\ 
35.0&(21,\, 28) && 11.8 & 3.7 & 5.4 & 10.1 &0.3 & 11.3 & 1.8 & 6.1
				   &6.9&0.8&8.7&0.5&5.4&3.7&1.2&--&\, 16.8 \\
35.2&(3.5,\, 35) && 0.5 & 1.9 & 4.0 & 6.2 &8.3 & 9.9 & 10.6 & 10.4
				   &9.4&7.7&5.7&3.7&2.0&0.8&0.1&--&\, 18.6 \\
70.1&(3.5,\, 70) && -- & 0.3 & 0.6 & 1.0 & 1.4 & 2.0 &2.6 & 3.1 & 3.7&4.2
			  &4.6&5.0&5.2&5.3&5.3&5.2 &\, 50.6 \\
105.1&(3.5,\, 105) && -- & -- & 0.2 &0.3 &0.5 &0.6 &0.8 & 1.1 &1.3&1.6
			   &1.8&2.1&2.3&2.6&2.8&3.0  &\, 79.0 \\
\hline\hline
\end{tabular}
\end{center}
\end{table*}

\begin{figure}
\begin{center}
\epsfig{file=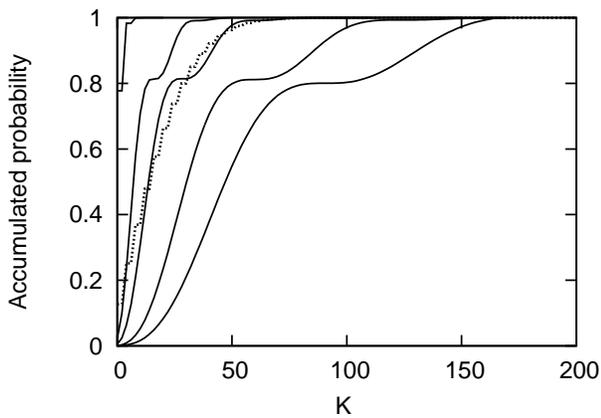,scale=1.3}
\caption{
Accumulated probability of finding the hyperspherical harmonics up to hypermomentum $K$ in the shifted 
Gaussian~(\ref{shifted.gauss}). Sets of 
$(\gamma_1, \gamma_2)$ shown are (3.5, 3.5), (3.5, 21), 
(3.5, 35), (25, 25), (3.5, 70), and (3.5, 105) in increasing order of $\sqrt{\gamma_1^2+\gamma_2^2}$. The dotted line denotes the case with (25, 25), and the solid lines the other cases. }
\label{kdistr}
\end{center}
\end{figure}

Table~\ref{HHexpansion} lists the values of $||f_K(s_1,s_2)||^2$ for 
some sets of $\gamma_1=\sqrt{\beta}s_1$ and $\gamma_2=\sqrt{\beta}s_2$ with $\beta=4\times 0.52$\,fm$^{-2}$.  
The rms radius, $\sqrt{\langle \rho^2\rangle}$, calculated with 
the wave function is approximately 
given by $\sqrt{\gamma_1^2+\gamma_2^2}/\sqrt{\beta}$. 
The value of $\gamma_1=3.5$ corresponds to 
the $\alpha \alpha$ distance, 
$\sqrt{2}s_1=\sqrt{2/\beta}\gamma_1\approx 3.43$\,
fm, which is on the order of the 
$\alpha \alpha$ distance of the $^8$Be($0^+$) resonance. The
$2\alpha$-$\alpha$ distance is given by $\sqrt{3/2\beta}\gamma_2$.  
The probability distribution quickly 
spreads to larger and more $K$ values as the system size becomes larger. 
For  $\sqrt{\gamma_1^2+\gamma_2^2}\approx 30$, corresponding to the rms radius of 21\,fm, the probability that exceeds $K=30$ already adds up to 10\%. In case where the rms radius exceeds  
70\,fm, the components with $K \leq 30$ are already smaller 
than 20\%.  Figure~\ref{kdistr} displays the  
accumulated probability, $\sum_{K'=0}^K ||f_{K'}(s_1,s_2) ||^2$, as a function of $K$. The plateau behavior seen in some curves is not a general feature but it is simply because $\gamma_1$ is set to 3.5. In fact 
the curve with $(\gamma_1, \gamma_2)=(25, 25)$ shows no  
plateau. 

Using the specific examples we have shown that the HH convergence turns out to be slower as 
the system size becomes larger. This property holds true 
in general. Suppose that for 
a given wave function $\Phi(x_1, x_2)$ depending on 
$(x_1, x_2)$ we want to approximate it in terms of a superposition of the  
HH functions. How many HH functions do we need? The values of the wave function at 
two points, $(x_1, x_2)$ and $(x_1', x_2')$, on the hypersphere of a radius $\rho$ differ when the 
distance of the two points is, say $D$, 
which is expressed as $D=2\rho |\sin \{(\alpha-\alpha')/2\}|
\approx \rho |\alpha-\alpha'|$, 
where $\alpha'$ is the hyperangle for $(x_1', x_2')$. Thus 
the number of mesh points needed to discretize $\alpha$ is on the order of 
\begin{align}
\frac{\pi}{2}\frac{1}{|\alpha-\alpha'|} \approx \frac{\pi}{2D}\rho.
\end{align}
This number corresponds to the number of needed HH functions since the hypermomentum $K$ is a quantum number related to the hyperangle $\alpha$. Therefore  
the above result clearly shows that an increasing number of the HH functions is needed as the system or $\rho$ increases.

\section{Partial wave contents of damped plane wave}
\label{pwc}

It is important to realize that the symmetrization in general brings
about high partial waves between $\alpha$ particles. 
Ogata {\it et al.}~\cite{ogata09} use the CDCC method in which 
only the $S$-wave continuum states of  
$2\alpha$ particles are discretized and in addition the symmetrization is neglected. We already point out 
in Sec.~\ref{discuss.lit} that the $D$-wave components are necessary 
to account for even the long-range coupling of the Coulomb potential. 
An interesting question is whether or not the $D$-wave components 
can be accounted for if the symmetrized basis is used in 
the CDCC calculation. 

To investigate this problem, 
we simulate the CDCC basis 
functions with a damped plane wave (DPW):
\begin{align}
&\Phi_{00}(k_1,k_2,\bm x)\notag \\
&=j_0(k_1x_1)j_0(k_2x_2)[Y_0(\hat{\bm
 x}_1)Y_0(\hat{\bm x}_2)]_{00}\exp\big(-\textstyle{\frac{1}{2}}a\rho^2\big). 
\label{dpw}
\end{align}
Here the relative motion corresponding to the coordinate $\bm x_1$ or 
$\bm x_2$ is basically free $S$-wave but its asymptotics 
is made to damp using the hyperscalar Gauss function. 
The parameter $a$ controls how far the DPW reaches. The 
wave numbers, $k_1$ and $k_2$, are parameters that determine the density 
of discretized states. The symmetrized DPW is obtained by 
\begin{align}
&\Psi_{00}(k_1,k_2)\notag \\
&=\Phi_{00}(k_1,k_2,\bm x)+
\Phi_{00}(k_1,k_2,\bm y)+\Phi_{00}(k_1,k_2,\bm z). 
\label{symm.dpw}
\end{align}

What differences do we have between 
the symmetrized DPW~(\ref{symm.dpw}) and 
the non-symmetrized DPW~(\ref{dpw}) in the 
continuum discretization at low energies?  Using a 
formula for the spherical Bessel function 
\begin{align}
&j_0(\sqrt{z^2+\zeta^2-2z \zeta\cos \theta})\notag \\
&=\sum_n(2n+1)j_n(z)j_n(\zeta)P_n(\cos \theta), 
\label{j0.exp} 
\end{align}
it is possible to expand the symmetrized DPW~(\ref{symm.dpw}) into partial waves as 
\begin{align}
&\Psi_{00}(k_1,k_2)\notag \\
&=
\sum_{\ell}f_{\ell}(k_1,k_2, x_1, x_2)[Y_{\ell}(\hat{\bm x}_1)Y_{\ell}(\hat{\bm x}_2)]_{00}
\exp\big(-\textstyle{\frac{1}{2}}a\rho^2\big),  
\label{symm.dpw.exp}
\end{align}
where $\ell$ is even and 
\begin{align}
&f_{\ell}(k_1,k_2, x_1, x_2)\notag \\
&=\delta_{\ell,0}j_0(k_1x_1)j_0(k_2x_2) +
\frac{1}{\sqrt{2\ell+1}}\sum_{nn'}\notag \\
& \times ((-1)^n+(-1)^{n'})(2n+1)(2n'+1)
\langle n0n'0|\ell 0\rangle^2 \notag \\
& \times  j_n(\textstyle{\frac{1}{2}}k_1x_1)j_{n'}(\textstyle{\frac{\sqrt{3}}{2}}k_2x_1)
j_{n}(\textstyle{\frac{\sqrt{3}}{2}}k_1x_2)j_{n'}(\textstyle{\frac{1}{2}}k_2x_2). 
\end{align}
The second term of the right-hand side of the above equation arises 
from the coordinate transformation from $\bm y$ and $\bm z$ to 
$\bm x$. 
Though Eq.~(\ref{dpw}) contains only the $S$-wave, its
symmetrization results in mixing higher partial waves. The quantity 
\begin{align}
W_{\ell}&=\int_0^{\infty}dx_1 x_1^2 \int_0^{\infty} dx_2 x_2^2
 \notag \\
&\times [f_{\ell}(k_1,k_2, x_1, x_2)]^2 \exp(-ax_1^2-ax_2^2) 
\end{align} 
gives the relative weight of $\ell$-wave contained in the symmetrized DPW.

To obtain an estimate of $W_{\ell}$, we choose a simple case, 
$k_2=0$.  
The rms radius of the DPW~(\ref{dpw}) with $k_2=0$ is given by 
$\sqrt{\langle {\cal M}_{00} \rangle}=\sqrt{C(\lambda)/a}$, 
where $\lambda=k_1^2/a$ and $C(\lambda)=2/3+\lambda/[3(e^\lambda-1)]$ is a monotone 
decreasing function bounded between $2/3<C(\lambda) \leq 1$. The kinetic energy with Eq.~(\ref{dpw}) is  given by 
$[C(\lambda)+\lambda/3]\varepsilon_0$, where $\varepsilon_0=3\hbar^2 a/2m_{\alpha}$ is the  
zero-point energy confined by the radius. Thus the kinetic energy approaches $\varepsilon=(\lambda/3)\varepsilon_0=\hbar^2k_1^2/2m_{\alpha}$ as $\lambda$ increases.  
Table~\ref{table.2} lists the probability, $W_{\ell}/\sum_{\ell}W_{\ell}$, of finding the component of 
$\ell$   
for some values of $\lambda$.  Only for $\lambda \leq 10$, 
the $\alpha \alpha$ relative motion is dominated by
$S$-wave, but otherwise the $S$-wave probability rapidly decreases to about 1/3. Suppose that the rms radius is of the order of 
$1/\sqrt{a}=500$ fm. 
If $\varepsilon$ is 0.1 MeV, $k_1^2/a$ turns out
to be about 4800 and the $S$-wave probability is
merely $1/3$. With $\varepsilon$ of 0.01 MeV, 
the $S$-wave probability is still 40\% at most. To have an $S$-wave dominance, 
$\varepsilon$ must reduce to 0.2 keV. 

The neglect of symmetrization leads to
an overestimation of low-partial wave contents. It appears that 
symmetrizing the non-symmetrized CDCC basis functions alone 
does not produce the $D$-wave components large enough to take care 
of the long-range Coulomb coupling.  

\begin{table}[b]
\begin{center}
\caption{Probability of finding the component with partial wave
 $\ell$ in the symmetrized damped plane wave~(\ref{symm.dpw}) with $k_2=0$ for several sets of $k_1^2/a$. Here 
$k_1$ is the wave number
 of the $\alpha \alpha$ relative motion and $a$ is the falloff parameter of
 the hyperscalar Gauss function. }  
\label{table.2}
\begin{tabular}{cccccccc}
\hline\hline
$k_1^2/a$ & & & & $\ell$& & & \\
\cline{2-8}
 & 0 & 2 & 4 & 6 & 8 & 10& $\geq$12\\
\hline
1 & 1.0 & 0.0 & 0.0 & 0.0 & 0.0 & 0.0 & 0.0\\
10 &0.907&0.092& 0.001 & 0.0 & 0.0 & 0.0 & 0.0\\
100 &0.404&0.255&0.214&0.095&0.026&0.005 & 0.001 \\
1000 &0.340&0.034&0.058&0.074&0.082&0.083 & 0.329 \\
10000 &0.334&0.004&0.006&0.009&0.012&0.014 & 0.621\\
\hline\hline
\end{tabular}
\end{center}
\end{table}

 \end{document}